\documentclass[preprint,aps,amsmath,showpacs,floatfix,draft]{revtex4}
\begin{document}

\title{Semiclassical coherent state propagator for systems with spin}

\author{A. D. Ribeiro$^{\dag\S}$, M. A. M. de Aguiar$^{\S}$ and
A F. R. de Toledo Piza$^{\dag}$}

\affiliation{$^{\dag}$ Instituto de F\'{\i}sica, Universidade de S\~ao Paulo,
USP 05315-970, S\~ao Paulo, S\~ao Paulo, Brazil}

\affiliation{$^{\S}$ Instituto de F\'{\i}sica ``Gleb Wataghin'',
Universidade Estadual de Campinas,
Unicamp 13083-970, Campinas, S\~ao Paulo, Brazil}

\begin{abstract}

We derive the semiclassical limit of the coherent state propagator
for systems with two degrees of freedom of which one degree of
freedom is canonical and the other a spin. Systems in this
category include those involving spin-orbit interactions and the
Jaynes-Cummings model in which a single electromagnetic mode
interacts with many independent two-level atoms. We construct a
path integral representation for the propagator of such systems
and derive its semiclassical limit. As special cases we consider
separable systems, the limit of very large spins and the case of
spin $1/2$.

\end{abstract}

%{03.65.Sq, 31.15.Gy}
% 03.65.Sq Semiclassical theories and applications
% 31.15.Gy Semiclassical methods
\maketitle

%%%%%%%%%%%%%%%%%%%%%%%%%%%%%%%%%%%%%%%%%%%%%%%%%%%%%%%
%%%%%%%%%%%%%%%%%%%%%%%%%%%%%%%%%%%%%%%%%%%%%%%%%%%%%%%
\section{Introduction}
%%%%%%%%%%%%%%%%%%%%%%%%%%%%%%%%%%%%%%%%%%%%%%%%%%%%%%%
%%%%%%%%%%%%%%%%%%%%%%%%%%%%%%%%%%%%%%%%%%%%%%%%%%%%%%%

The spin-orbit interaction plays important roles in many areas of
physics, from atomic physics to condensed matter. The quantum
description of systems with such interactions requires the use of
Hilbert spaces which are the direct product of the orbital space
(for which the coordinate or the momentum eigenstates form a
basis) times the intrinsic space of the spin.

In quantum optics a similar situation arises in the study of the
interaction between atoms and electromagnetic modes in a cavity.
When only two states of the atoms are relevant, as for example in
the ammonia maser, these two states can be formally represented by
a spin $1/2$. The state of a set of $N$ such atoms can be likewise
represented by the states of an angular momentum of larger
magnitude. The Hamiltonian describing their interaction with a
single electromagnetic mode of a cavity can therefore be written
in terms of the operators $\hat{a}$ and $\hat{a}^{\dagger}$, which
annihilate and create excitations of the quantized electromagnetic
mode, and $\hat{J}_z$, $\hat{J}_+$ and $\hat{J}_-$, of the angular
momentum.

The semiclassical behavior of such systems has drawn attention for
quite a long time. One natural representation for the study of
this limit is that of coherent states. The semiclassical limit of
the coherent state propagator for both the Weyl and the SU(2)
group has already been studied in detail. The purpose of this
paper is to derive the semiclassical limit of the coherent state
propagator for general systems with two degrees of freedom in
which one degree is canonical and the other a spin.

The quantum propagator $K(b''^*,b',T) \equiv \langle b''|
e^{-i\hat{H}T/\hbar}| b'\rangle$ represents the probability
amplitude that the initial state $|b'\rangle$ be measured as
$|b''\rangle$, after a time $T$, when evolved by the Hamiltonian
$\hat{H}$. The propagator is the essential ingredient for quantum
dynamical calculations and it is also fundamental in the study of
the quantum-classical connection. Semiclassical approximations of
the propagator in the coordinates and momentum representations
were studied initially by Feynman \cite{feynman} and later by many
others \cite{books}. Semiclassical formulae for the propagator in
the basis of coherent states appeared for the first time in the
works of Klauder \cite{klauder78,klauder79}. Although these papers
have treated the propagator for both canonical coherent states
$|z\rangle$ and spin coherent states $|s\rangle$, not much work
has been done on these two basis simultaneously. The development
of the semiclassical theory occurred independently on each
representation.

For canonical coherent states, Weissman \cite{weis82,weis83}
re-derived the results of Klauder using the semiclassical theory
of Miller \cite{miller}. In Weissman's work, and also in the
original Klauder's papers, the fluctuations around the critical
trajectory have not been accurately performed, and a `phase'
factor was missed. In spite of this, a first numerical application
of the semiclassical formula was performed by Adachi \cite{adachi}
for an two-dimensional chaotic map, obtaining good agreement with
exact quantum results. The correct evaluation of the second order
fluctuations appeared in the works of Baranger and Aguiar
\cite{bar0}, Xavier and Aguiar \cite{xav3,xav1,xav2} and,
independently, by Kochetov \cite{kochetov98}. However, a detailed
derivation of the semiclassical coherent state propagator for
one-dimensional systems has appeared only later in \cite{bar}.
Extensions of the formula for two dimensional degrees of freedom
and applications to chaotic systems have been performed in Ref.
\cite{rib1}.

There are two differences between the semiclassical formula of
Baranger {\it et al} and that of Klauder and Weissman. First,
there is the extra `phase' in the new formula, which is usually
complex and, in fact, is a signature of the basis of coherent
states. It is related to the overcompletness of the basis, since
changing the resolution of the unity also changes this term
\cite{bar,coe}. A second difference consists in replacing the Weyl
symbol $H$ of the Hamiltonian operator $\hat{H}$ by the average
$\tilde{H}_z\equiv\langle z | \hat{H}|z\rangle$. This implies that
the classical trajectories entering in the formula are subjected
to $\tilde{H}_z$ instead of $H$. The dynamics with $\tilde{H}$
actually appears naturally in the work by Klauder, but it was
changed back to $H$ to be consistent with the lack of the extra
exponential factor of the formula. As discussed in Ref.
\cite{bar}, both these changes are essential to get good agreement
between quantum and semiclassical results. We note that the
semiclassical coherent state propagator also presents
singularities and discontinuities due to phase space caustics and
the Stokes phenomenon \cite{adachi,rubin,rib1,aguiarjpa,rib3}.

Approximations for spin coherent states have also been studied by
Kuratsuji and Suzuki \cite{kura1}. The basic difference between
their approach and Klauder's is that the later represents the
classical trajectories in a Bloch sphere by means of angle
variables while Kuratsuji and Suzuki represents the same dynamics
in terms of an stereographic projection of the Bloch sphere on a
complex plane. As is the case of canonical coherent states, a
detailed derivation in the spin coherent state propagator only
appeared later with Solari \cite{sol}, where features similar to
those appearing in the canonical case were found: an extra
exponential term, and the underlying classical dynamics governed
by the average Hamiltonian $\tilde{H}_s \equiv \langle
s|\hat{H}|s\rangle$. Vieira and Sacramento \cite{vieira} and
Kochetov \cite{kochetov} have also derived the same formula
independently. Yet another detailed derivation has also been
presented by Stone {\em et al} \cite{stone}, focusing on the
importance of the extra term (see also Ref. \cite{plet}), which
has received the name of {\em Solari-Kochetov phase}. The
singularities and discontinuities of the semiclassical spin
propagator are discussed in Refs. \cite{garg,marcel1}.

In spite of the extensive work on the semiclassical theory for the
canonical and spin representations, there are very few results
considering the two bases simultaneously. Alscher and Grabert
\cite{alscher} have calculated the semiclassical coherent state
propagator for the spin $1/2$ Jaynes-Cummings model showing that
it corresponds to the exact quantum result. Pletyukhov et al
\cite{plet2002,plet2003} derived a semiclassical trace formula for
systems with spin, but did not consider the semiclassical
propagator itself in detail. An overview of the semiclassical
approaches for spin-orbit interactions and associated trace
formulae was recently published by Amann and Brack \cite{amann},
and a general semiclassical theory for Hamiltonians which are
linear in spin operators has also been formulated
\cite{plet2002,plet2003,zaitsev1,zaitsev2}.

In this paper we derive a semiclassical formula for general
canonical-spin systems using path integrals. We show that our
formula agrees with the previous results
\cite{bar,sol,vieira,kochetov} in the separable case (no
spin-orbit interaction) and that it reduces to the 2-D canonical
coherent state propagator in the limit of very large spins.
Finally we discuss the limit of validity of the approximation for
spin-$\frac{1}{2}$ systems.

\section{Basic definitions}
%%%%%%%%%%%%%%%%%%%%%%%%%%%%%%%%%%%%%%%%%%%%%%%%%%%%%%%
%%%%%%%%%%%%%%%%%%%%%%%%%%%%%%%%%%%%%%%%%%%%%%%%%%%%%%%

The propagator in the canonical-spin representation is given by
\begin{equation}
K \left( z''^*, s''^*, z', s', T \right) = \langle  z'' , s'' |
e^{-i\hat{H}T/\hbar} |  z' , s'\rangle , \label{QP}
\end{equation}
where $|  z , s\rangle \equiv |  z\rangle \otimes | s \rangle $ is
the product of a canonical coherent state $ | z \rangle $ and a
spin coherent state $| s \rangle$. While $ | z \rangle $ is
defined in the `particle' Hilbert space,  $| s \rangle$ is defined
in the $(2j+1)$-dimensional space of an angular momentum $j$.
These states can be written as
\begin{equation}
|z \rangle =   e^{z  \hat{a}^{\dagger} -\frac{1}{2} |z|^2} |0
\rangle \qquad \mathrm {and} \qquad |s \rangle = \frac{e^{s \hat{J}_+}} {\left(
1+ |s|^2 \right)^j}  | -j \rangle , \label{CSdef}
\end{equation}
where $z$ and $s$ are complex numbers, $\hat{a}^{\dagger}$ is the
canonical creator operator, $\hat{J}_+$ is the raising spin
operator, $|0 \rangle$ is the harmonic oscillator vacuum state and
$| -j \rangle$ is the extremal eigenstate of $\hat{J}_3$ with
eingenvalue $-j$. The coherent states are non-orthogonal with
\begin{equation}
\langle z_1 | z_2 \rangle = \exp{ \left\{- \frac{1}{2} |z_1|^2 +
z_1^* z_2 - \frac{1}{2}|z_2|^2 \right\}} \qquad \mathrm{and}
\qquad \langle s_1 | s_2 \rangle = \frac{\left( 1 + s_1^* s_2
\right)^{2j}}{\left( 1 + |s_1 |^2 \right)^{j} \left( 1 + |s_2 |^2
\right)^{j}} , \label{NO}
\end{equation}
and satisfy
\begin{eqnarray}
\int \frac{\mathrm{d} x \, \mathrm{d} y}{\pi}|z \rangle \langle z
| \equiv 1_{(z)} \qquad \mathrm{and} \qquad \frac{2j+1}{\pi}\int
\frac{\mathrm{d}X \, \mathrm{d}Y }{\left( 1 + |s |^2
\right)^{2}}|s \rangle \langle s | \equiv 1_{(s)} , \label{OC}
\end{eqnarray}
where $x$ and $y$ are the real and imaginary parts of $z$ and $X$
and $Y$ the real and imaginary parts of $s$. Throughout this paper
we shall use lower case letters to refer to the canonical
variables and corresponding upper case letters to refer to the
spin variables.

Finally, the complex number $z$ labelling the canonical coherent
state can be written explicitly in terms of position and momentum
variables as
\begin{equation}
z = \frac{1}{\sqrt{2}}\left(\frac{q}{b} + i\frac{p}{c}\right)
\label{zqp}
\end{equation}
where $q$ and $p$ are the average values of the position and
momentum operators, respectively, and the variances $b$ and $c$
satisfy $bc = \hbar$.

%%%%%%%%%%%%%%%%%%%%%%%%%%%%%%%%%%%%%%%%%%%%%%%%%%%%%%
%%%%%%%%%%%%%%%%%%%%%%%%%%%%%%%%%%%%%%%%%%%%%%%%%%%%%%
\section{Path Integral Formulation}

In this section we construct a path integral representation for
the quantum propagator~(\ref{QP}). As usual we divide the time $T$
into $N$ small intervals of size $\epsilon=T/N$ and insert
resolutions of unity between each propagation step. We obtain
\begin{eqnarray}
K \left( z''^*, s''^*, z', s', T \right) &=&  \lim_{\epsilon\rightarrow 0}
  \int
  \prod_{k=1}^{N-1}
        \left\{ \left(\frac{2j+1}{\pi^2}\right)
        \frac{\mathrm{d} x_k \mathrm{d}y_k
        \mathrm{d} X_k \mathrm{d} Y_k}
        {\left( 1 + |s_k |^2 \right)^{2}} \right\}
  \nonumber \\
  &\times&
  \prod_{k=0}^{N-1}
         \langle z_{k+1},s_{k+1}|e^{-i\hat{H}\epsilon/\hbar}| z_k , s_k \rangle  ,
\label{DPI}
\end{eqnarray}
where we define $| z_0 , s_0 \rangle \equiv | z' , s' \rangle$ and
$\langle z_N , s_N | \equiv \langle z'' , s'' |$. In the limit
$\epsilon \rightarrow 0$ the infinitesimal propagators can be
written as
\begin{equation}
\langle z_{k+1},s_{k+1}|e^{-i\hat{H}\epsilon/\hbar}| z_k , s_k
\rangle =
   \langle z_{k+1},s_{k+1}| z_k , s_k \rangle
     \exp{\left\{ -\frac{i}{\hbar} \epsilon \tilde{H}_{k+1/2} \right\}},
     \label{eq8}
\end{equation}
where
\begin{equation}
\tilde{H}_{k+1/2} \equiv
     \frac{ \langle z_{k+1},s_{k+1}| \hat{H} | z_k , s_k \rangle}
     {\langle z_{k+1},s_{k+1}| z_k , s_k \rangle}
\end{equation}
and
\begin{equation}
\langle z_{k+1},s_{k+1}| z_k , s_k \rangle =
   \frac{\left( 1 + s_{k+1}^* s_{k} \right)^{2j}}
       {\left( 1 + |s_{k+1} |^2 \right)^{j}
       \left( 1 + |s_{k} |^2 \right)^{j}}
   \exp{ \left\{
       - \frac{1}{2} |z_{k+1}|^2 + z_{k+1}^* z_{k}
       - \frac{1}{2}|z_{k}|^2 \right\}}.
\end{equation}
With these considerations the propagator can be written as
\begin{eqnarray}
K \left( z''^*,  s''^*, z', s', T \right) =
  \lim_{\epsilon\rightarrow 0} \left(\frac{2j+1}{\pi^2}\right)^{N-1}
  \int
  \prod_{k=1}^{N-1} \left\{\mathrm{d} x_k \mathrm{d}y_k
        \mathrm{d} X_k \mathrm{d} Y_k
  \right\} e^{F},
\label{DPIs1}
\end{eqnarray}
where
\begin{equation}
F = F_z + F_s -\frac{i}{\hbar} \sum_{k=0}^{N-1}
\epsilon\tilde{H}_{k+1/2} ,
\label{F}
\end{equation}
with

\begin{equation}
F_z  =
  \frac{i}{\hbar} \sum_{k=0}^{N-1}
     \frac{i\hbar}{2}
     \left[ z_{k} z_k^* - 2 z_{k+1}^* z_k + z_{k+1}^* z_{k+1} \right]
\end{equation}
and
\begin{equation}
F_s =
  \frac{i}{\hbar} \left\{
  -i\hbar j  \sum_{k=0}^{N-1}
  \ln
     \left[
            \frac{\left( 1 + s_{k+1}^* s_k \right)^{2}}
            {\left( 1 + s_k^* s_k \right)
             \left( 1 + s_{k+1}^* s_{k+1} \right)}
     \right]
     -2i\hbar  \sum_{k=1}^{N-1}
            \ln \left[ \frac{1}{( 1 + s_k^* s_k )} \right] \right\}  .
\end{equation}
Eq. (\ref{DPIs1}) is a discretized path integral representation of
the propagator (\ref{QP}). In the following sections we shall
consider the formal semiclassical limit $\hbar \rightarrow 0 $ and
$j \rightarrow \infty$, keeping the product $j \hbar \equiv J$
constant.

%%%%%%%%%%%%%%%%%%%%%%%%%%%%%%%%%%%%%%%%%%%%%%%%%%%%%%
%%%%%%%%%%%%%%%%%%%%%%%%%%%%%%%%%%%%%%%%%%%%%%%%%%%%%%
\section{The Semiclassical Limit }

In the semiclassical limit, the integrals in Eq. (\ref{DPIs1}) can
be performed by the saddle point method. The method consists
basically in approximating the exponent $F$ by a quadratic form
and performing the resulting Gaussian integrals. The quadratic
form is obtained by expanding $F$ around its critical points. In
the next subsections we shall: (a) find the critical points of
$F$, and therefore the critical path; (b) compute $F$ at the
critical path; (c) expand $F$ to second order around the critical
path and compute the Gaussian integrals.

%%%%%%%%%%%%%%%%%%%%%%%%%%%%%%%%%%%%%%%%%%%%%%%%%%%%%%
\subsection{The Critical Path }

We begin by looking for critical
points of $F$. They satisfy the equations
\begin{eqnarray}
\frac{\partial F}{\partial z_m} =
\frac{\partial F}{\partial z_m^*} =
\frac{\partial F}{\partial s_m} =
\frac{\partial F}{\partial s_m^*} =0, \qquad
m=1,\ldots,N-1 .
\end{eqnarray}
As $\tilde{H}_{k+1/2} = \tilde{H}_{k+1/2} (z_{k+1}^*, s_{k+1}^*,
z_{k},s_{k})$, these equations can be explicitly written as
\begin{eqnarray}
\begin{array}{c}
\displaystyle{\frac{i\epsilon}{\hbar} \frac{\partial \tilde{H}_{m+1/2}}{\partial
z_m} =
   z_{m+1}^* - z_{m}^* } ,\\
\displaystyle{\frac{i\epsilon}{\hbar} \frac{\partial \tilde{H}_{m-1/2}}{\partial
z_m^*} =
   z_{m-1} - z_{m}},
\end{array} \label{DF2l}
\end{eqnarray}
and
\begin{eqnarray}
\begin{array}{l}
\displaystyle{\frac{i\epsilon}{2\hbar} \frac{\partial
\tilde{H}_{m+1/2}}{\partial s_m} =
   j
      \left\{
          \frac{s_{m+1}^*}{1+s_{m+1}^*s_m} - \frac{s_{m}^*}{1+s_{m}^*s_m}
      \right\}
      -  \frac{s_{m}^*}{1+s_{m}^*s_m}} , \\
\displaystyle{\frac{i\epsilon}{2\hbar} \frac{\partial
\tilde{H}_{m-1/2}}{\partial s_m^*} =
   - j
      \left\{
        \frac{s_{m}}{1+s_{m}^*s_m} - \frac{s_{m-1}}{1+s_{m}^*s_{m-1}}
      \right\}
      - \frac{s_{m}}{1+s_{m}^*s_m}}.
\end{array}\label{DF4l}
\end{eqnarray}
It is important to emphasize that, because $m=1,\ldots,N-1$, the
variables $z_0^*$, $s_0^*$, $z_N$, $s_N$ do not enter in Eqs.
(\ref{DF2l}) and (\ref{DF4l}): the critical path, defined by the
set of critical points, depends only on $z_0$, $s_0$, $z_N^*$ and
$s_N^*$, and not on $z_0^*$, $s_0^*$, $z_N$, $s_N$.

In the limit $\epsilon \rightarrow 0$ Eqs. (\ref{DF2l}) become

\begin{eqnarray}
\frac{i}{\hbar} \frac{\partial \tilde{H}}{\partial z} = \dot{z}^*
\qquad \mathrm{and} \qquad \frac{i}{\hbar} \frac{\partial
\tilde{H}}{\partial z^*} = - \dot{z} \label{ehz}
\end{eqnarray}
where $\tilde{H} \equiv \langle z,s |\hat{H}| z,s \rangle$. In
terms of $q$ and $p$ (see Eq.(\ref{zqp})) this corresponds to the
usual Hamilton's equations
\begin{eqnarray}
\frac{\partial \tilde{H}}{\partial p} = \dot{q} \qquad
\mathrm{and} \qquad \frac{\partial \tilde{H}}{\partial q} = -
\dot{p} \label{ehzqp}
\end{eqnarray}

For Eqs. (\ref{DF4l}) the calculation is slightly more involved
but the result is also very simple. We obtain, in the limit
$\epsilon \rightarrow 0$ and $\hbar \rightarrow 0$ with $j = J /
\hbar$,
\begin{eqnarray}
\frac{\partial \tilde{H}}{\partial s} =
           - \frac{2i J \dot{s}^*}{\left( 1+s^*s \right)^2}
\qquad \mathrm{and} \qquad
 \frac{\partial \tilde{H}}{\partial s^*} =
        \frac{2i J \dot{s}}{\left( 1+s^*s \right)^2}. \label{ehs}
\end{eqnarray}

The trajectories described by Eqs. (\ref{ehz}) and (\ref{ehs})
define the critical path of the Feynman integral (\ref{DPIs1}).
Nevertheless, these trajectories must satisfy the boundary
conditions $z(0)=z'$, $s(0)=s'$, $z^*(T)=z''^*$ and
$s^*(T)=s''^*$, as can be seen from Eqs. (\ref{DF2l}) and
(\ref{DF4l}). As discussed in detail in \cite{bar}, these
trajectories are usually complex and the variables $z$ and $z^*$
are not generally the complex conjugate of each other, the same
happening between $s$ and $s^*$. Therefore, it is convenient to
set a new notation
\begin{eqnarray}
z=u ,\qquad s=U, \qquad z^*=v \qquad \mathrm{and} \qquad s^*=V .
\label{nn}
\end{eqnarray}
In terms of these new variables, the equations of motion
(\ref{ehz}) and (\ref{ehs}) become
\begin{eqnarray}
\frac{i}{\hbar} \frac{\partial \tilde{H}}{\partial u} = \dot{v},
\qquad \frac{i}{\hbar} \frac{\partial \tilde{H}}{\partial v} = -
\dot{u}, \qquad \frac{i}{\hbar}
   \frac{\partial \tilde{H}}{\partial U} =
           \frac{ 2j \dot{V}}{\left( 1+UV
           \right)^2},   \qquad \frac{i}{\hbar}
   \frac{\partial \tilde{H}}{\partial V} =
        \frac{-2j\dot{U}}{\left( 1+UV \right)^2} ,\label{emnn}
\end{eqnarray}
with boundary conditions
\begin{eqnarray}
u(0)=z', \qquad U(0)=s', \qquad v(T)=z''^*, \qquad V(T)=s''^*.
\label{bbnn}
\end{eqnarray}
Since $z'^*$, $s'^*$, $z''$ and $s''$ do not appear in the
boundary conditions, the value of $u(T)$, $U(T)$, $v(0)$ and
$V(0)$ are determined by the integration of Eqs. (\ref{emnn}).
From now on we shall use this new notation to refer to the complex
classical trajectories. The discrete variables $z_m$, $s_m$,
$z_m^*$ and $s_m^*$ shall also be replaced by $u^m$, $U^m$, $v^m$
and $V^m$, respectively.

%%%%%%%%%%%%%%%%%%%%%%%%%%%%%%%%%%%%%%%%%%%%%%%%%%%%%%
\subsection{The Complex Action}

The function $F$ appearing in Eq. (\ref{F}) can now be calculated
at the classical trajectory. We use a bar over the variables
 to indicate that they are calculated at the critical path.
We have
\begin{eqnarray}
F &=&
  \frac{i}{\hbar} \sum_{k=0}^{N-1}
  \left\{
     \frac{i\hbar}{2}
     \left[ \bar u^{k} \bar v^k - 2 \bar v^{k+1}
     \bar u^k + \bar v^{k+1} \bar u^{k+1} \right]
     -i \hbar j \ln
     \left[
       \frac{\left( 1 + \bar V^{k+1} \bar U^k \right)^2}
       {\left( 1 + \bar V^k \bar U^k \right)
       \left( 1 + \bar V^{k+1} \bar U^{k+1} \right)}
     \right]
  \right\} \nonumber \\
  &-&
  \frac{i}{\hbar} \sum_{k=0}^{N-1} \epsilon
  \tilde{H}_{k+1/2} (\bar{v}^{k+1},\bar{V}^{k+1},\bar{u}^{k},\bar{U}^{k} )
  -2 \sum_{k=1}^{N-1} \ln \left( 1 + \bar V^k \bar U^k \right)
  \nonumber \\
  &-&
  \frac{1}{2} \left( z'z'^* +z''z''^*-\bar u^{0} \bar v^0 -
  \bar u^{N} \bar v^{N}
  \right)
   -j \ln \left[ \frac{(1+s's'^*)(1+s''s''^*)}{(1+\bar U^0\bar V^0)
   (1+\bar U^N\bar V^{N})}
   \right]. \nonumber
   \label{eq29}
\end{eqnarray}
As usual we have replaced $z'^*,s'^*,z'',s''$ by $\bar v^0,\bar
V^0,\bar u^N,\bar v^N$ in the first line and corrected for this in
the last line. Taking the limit $\epsilon\rightarrow 0$ we find,
after some algebra,
\begin{align}
F =
  \frac{i}{\hbar} \mathcal{S}(z''^*,s''^*,z',s',T)
   - \Lambda - 2 \sum_{k=1}^{N-1}
            \ln \left( 1 + \bar V^k \bar U^k \right),
            \label{Fzero}
\end{align}
where $\mathcal{S}(z''^*,s''^*,z',s',T)$ is the complex action
\begin{align}
\mathcal{S}(z''^*,s''^*,z',s',T) &=
     \int_0^T
     \left\{
     \frac{i\hbar}{2} \left(\dot{\bar u} \bar v -
     \dot{\bar v} \bar u \right) -i\hbar j
     \left(  \frac{\bar U \dot{\bar V} -
     \bar V \dot{\bar U}}
     { 1 + \bar U \bar V }  \right)  - {\tilde H}
     \right\}  dt  \nonumber \\
        & -\frac{i\hbar}{2}\left(
        \bar u' \bar v' + \bar u'' \bar v'' \right)
    -i\hbar j \ln \left[
   (1+\bar U'\bar V')(1+\bar U''\bar V'') \right]
   \label{action}
\end{align}
and
\begin{equation}
\Lambda = \frac{1}{2}\left( |z'|^2 + |z''|^2\right) + j \ln
\left[(1+|s'|^2)(1+|s''|^2) \right] \label{lambda}
\end{equation}
is a `normalization term'. The limit $\epsilon \rightarrow 0$ has
not been taken on last term of Eq.(\ref{Fzero}) because this term
is going to get cancelled later on.

It can be checked that
\begin{eqnarray}
\frac{\partial \mathcal{S}}{\partial \bar u'} = -i\hbar \bar v',
\qquad \frac{\partial \mathcal{S}}{\partial \bar v''} = -i\hbar
\bar u'', \label{dif1}
\end{eqnarray}
\begin{eqnarray}
\frac{\partial \mathcal{S}}{\partial \bar U'} = \frac{-2i\hbar
j\bar V'}{1+\bar U'\bar V'}, \qquad \frac{\partial
\mathcal{S}}{\partial \bar V''} = \frac{-2i\hbar j \bar
U''}{1+\bar V''\bar U''}, \label{dif2}
\end{eqnarray}
and
\begin{eqnarray}
\frac{\partial \mathcal{S}}{\partial T} = -{\tilde H},
\label{variaS}
\end{eqnarray}
where a single (double) prime to denotes initial (final) time.

%%%%%%%%%%%%%%%%%%%%%%%%%%%%%%%%%%%%%%%%%%%%%%%%%%%%%%
\subsection{Fluctuations around the Critical Path}

In the semiclassical limit, the only relevant points in the path
integral of Eq. (\ref{DPIs1}) are the saddle points. In the
present case, they define trajectories governed by Eqs.
(\ref{emnn}) [or by their discretized forms, Eqs. (\ref{DF2l}) and
(\ref{DF4l})]. The exponent $F$ in Eq. (\ref{DPIs1}) has to be
integrated over the intermediate points $\mathbf{x} \equiv (u^1,
U^1, v^1, V^1, \ldots, u^{N-1}, U^{N-1},v^{N-1},V^{N-1})^T$.
Expanding $F$ around the critical trajectory $\bar{\mathbf{x}}$ we
obtain
\begin{eqnarray}
F(\mathbf{x}) =  F(\bar{\mathbf{x}}) - \frac{1}{2}
           \delta\mathbf{x}^T [-\delta^2 F(\bar{\mathbf{x}})]
           \delta\mathbf{x},
\end{eqnarray}
where  $\delta\mathbf{x}\equiv \mathbf{x}-\bar{\mathbf{x}}$. The
matrix $[-\delta^2 F(\bar{\mathbf{x}})]$ contains  the second
derivatives of $F$ calculated with the critical trajectory.
Substituting this result in Eq. (\ref{DPIs1}) and considering the
jacobian
\begin{eqnarray}
\left\{ \mathrm{d}  x_k \mathrm{d}  y_k \mathrm{d}
X_k \mathrm{d}  Y_k \right\} \rightarrow -\frac{1}{4}
\left\{ \mathrm{d} z_k \mathrm{d} z_k^* \mathrm{d}  s_k \mathrm{d}
s_k^* \right\} \equiv -\frac{1}{4} \left\{ \mathrm{d}  u^k
\mathrm{d} v^k\mathrm{d}  U^k \mathrm{d}  V^k \right\} \nonumber ,
\end{eqnarray}
we find
\begin{eqnarray}
K  = e^{F(\bar{\mathbf{x}})}
  \lim_{\epsilon\rightarrow 0}
  \left\{
   \left(\frac{2j+1}{-4\pi^2}\right)^{N-1}
  \int
  \prod_{k=1}^{N-1} \left\{
  \mathrm{d} [\delta u^k] \mathrm{d} [\delta v^k]
  \mathrm{d} [\delta U^k] \mathrm{d} [\delta V^k]  \right\}
  e^{-\frac{1}{2} \delta \mathbf{x}^T [-\delta^2 F(\bar{\mathbf{x}})]
  \delta \mathbf{x} }
  \right\},
\label{eqexp1}
\end{eqnarray}
where $F(\bar{\mathbf{x}})$ is given by Eq. (\ref{Fzero}).
The matrix $[-\delta^2 F(\bar{\mathbf{x}})]$ is written as
{
\begin{eqnarray}
%[-\delta^2 F(\bar{\mathbf{x}})]=
\left(
\begin{array}{cccc|cccc|c}
%%%%%%%%%%%%%%%%%%%%%%%%%%%%%%%%%%%%%%%%%%%%%%%%linha1
\mathcal{H}_{11}^{N-1} & \mathcal{H}_{21}^{N-1} & 1 & 0
&  0                     & 0  & 0    &  0      &  \ldots  \\
%%%%%%%%%%%%%%%%%%%%%%%%%%%%%%%%%%%%%%%%%%%%%%%%linha2
\mathcal{H}_{21}^{N-1}      &  \mathcal{H}_{22}^{N-1}     & 0 &
{B}^{N-2}    & 0 & 0 & 0   & 0  & \ldots \\
%%%%%%%%%%%%%%%%%%%%%%%%%%%%%%%%%%%%%%%%%%%%%%%%linha3
1  & 0  & \mathcal{H}_{33}^{N-2} & \mathcal{H}_{43}^{N-2} &
\mathcal{H}_{13}^{N-2}-1 & \mathcal{H}_{23}^{N-2}    & 0
& 0  & \ldots  \\
%%%%%%%%%%%%%%%%%%%%%%%%%%%%%%%%%%%%%%%%%%%%%%%%linha4
0  & {B}^{N-2} & \mathcal{H}_{43}^{N-2} & \mathcal{H}_{44}^{N-2} &
\mathcal{H}_{41}^{N-2} & \mathcal{H}_{24}^{N-2}-\mathcal{B}^{N-2}
& 0 & 0 &\ldots       \\ \hline
%%%%%%%%%%%%%%%%%%%%%%%%%%%%%%%%%%%%%%%%%%%%%%%%linha5
0   & 0  & \mathcal{H}_{31}^{N-2}-1 & \mathcal{H}_{41}^{N-2} &
\mathcal{H}_{11}^{N-2}   & \mathcal{H}_{21}^{N-2}       & 1
& 0  & \ldots \\
%%%%%%%%%%%%%%%%%%%%%%%%%%%%%%%%%%%%%%%%%%%%%%%%linha6
0  & 0 & \mathcal{H}_{23}^{N-2} &
\mathcal{H}_{42}^{N-2}-\mathcal{B}^{N-2}   &
\mathcal{H}_{21}^{N-2}  & \mathcal{H}_{22}^{N-2}  & 0       &
{B}^{N-3}   & \ldots \\
%%%%%%%%%%%%%%%%%%%%%%%%%%%%%%%%%%%%%%%%%%%%%%%%linha7
0 & 0  & 0  & 0 & 1 & 0 & \mathcal{H}_{33}^{N-3} &
\mathcal{H}_{43}^{N-3}  & \ldots        \\
%%%%%%%%%%%%%%%%%%%%%%%%%%%%%%%%%%%%%%%%%%%%%%%%linha8
0 & 0 & 0 & 0 & 0 & {B}^{N-3} & \mathcal{H}_{43}^{N-3} &
\mathcal{H}_{44}^{N-3} & \ldots     \\
\hline
%%%%%%%%%%%%%%%%%%%%%%%%%%%%%%%%%%%%%%%%%%%%%%%%linha9
\vdots & \vdots &\vdots &\vdots &\vdots &\vdots
&\vdots &\vdots & \ddots \\
\end{array}\right) ,\nonumber
\end{eqnarray}}
where
\begin{equation}
B^m= \frac{2(j+1)}{(1+\bar V^m \bar U^m)^2} \quad, \qquad  \qquad
\mathcal{B}^m = \frac{2j}{(1+\bar V^{m+1} \bar U^m)^2}
\end{equation}
and $\mathcal{H}_{ij}^{m}\equiv\partial^2 {\tilde{\cal
H}}_{m+1/2}/\partial \chi_i^{m_i} \partial \chi_j^{m_j}$.
In this last definition we are using the compact notation

\begin{eqnarray}
\chi_1\equiv u, \qquad \chi_2 \equiv U, \qquad \chi_3\equiv v
\qquad \mathrm{and} \qquad  \chi_4\equiv V. \label{compactn}
\end{eqnarray}
In addition, $m_i$ ($m_j$) equals to $m+1$ when $i$ ($j$) is equal
to 3 or 4, and equals to $m$ when $i$ ($j$) is equal to 1 or 2. In
order to cancel the last term in Eq.(\ref{Fzero}) and also the
factor $(2j+1)^{N-1}$ in the pre-factor of Eq.(\ref{eqexp1}) we
change the variables associated with the spin by the
transformation $\delta U^m = \delta \tilde{U}^m /\sqrt{B^{m}}$ and
$\delta V^m = \delta \tilde{V}^m /\sqrt{B^{m}}$. Eq.
(\ref{eqexp1}) then becomes

\begin{eqnarray}
K  = e^{\frac{i}{\hbar}\mathcal{S}(\bar{\mathbf{x}})-\Lambda}
  \lim_{\epsilon\rightarrow 0}
  \left\{
   \left(\frac{-1}{4\pi^2}\right)^{N-1}
  \int
  \prod_{k=1}^{N-1} \left\{
  \mathrm{d} [\delta u^k] \mathrm{d} [\delta v^k]
  \mathrm{d} [\delta \tilde U^k] \mathrm{d} [\delta \tilde V^k]  \right\}
  e^{-\frac{1}{2} \delta \tilde{ \mathbf{x}}^T [-\delta^2 \tilde F(\bar{\mathbf{x}})]
  \delta \tilde{\mathbf{x}} }
  \right\},
\label{eqexp2}
\end{eqnarray}
where $\mathcal S (\bar{\mathbf x})$ and $\Lambda$ are given by
Eqs. (\ref{action}) and (\ref{lambda}) respectively and
$\delta\tilde{\mathbf{x}} \equiv (\delta u^1,  \delta \tilde U^1,
\delta v^1, \delta \tilde V^1,\ldots, \delta u^{N-1}, \delta
\tilde U^{N-1},\delta v^{N-1}, \delta \tilde V^{N-1})^T$ and
$[-\delta^2 \tilde F(\bar{\mathbf{x}})]$ is the matrix

\begin{eqnarray}
-\delta^2 \tilde{F}(\bar{\mathbf{x}})=
\left(
\begin{array}{cccc}
%%%%%%%%%%%%%%%%%%%%%%%%%%%%%%%%%%%%%%%%%%%%%%%%linha1
\mathrm{W}^{N-1}  &  \mathrm{R}^{N-1} & 0 & \ldots   \\
%%%%%%%%%%%%%%%%%%%%%%%%%%%%%%%%%%%%%%%%%%%%%%%%linha2
\left(\mathrm{R}^{N-1}\right)^T & \mathrm{W}^{N-2} & \mathrm{R}^{N-2} &  \ldots \\
%%%%%%%%%%%%%%%%%%%%%%%%%%%%%%%%%%%%%%%%%%%%%%%%linha3
0 & \left(\mathrm{R}^{N-2}\right)^T & \mathrm{W}^{N-3} & \ldots \\
%%%%%%%%%%%%%%%%%%%%%%%%%%%%%%%%%%%%%%%%%%%%%%%%linha4
\vdots &\vdots &\vdots & \ddots \\
\end{array}\right), \label{matilde}
\end{eqnarray}
where
\begin{eqnarray}
\mathrm{W}^{k}&=& \left(
\begin{array}{cccc}
%%%%%%%%%%%%%%%%%%%%%%%%%%%%%%%%%%%%%%%%%%%%%%%%linha1
\mathcal{H}_{11}^{k} & b^{k} \mathcal{H}_{21}^{k} & 1 & 0  \\
%%%%%%%%%%%%%%%%%%%%%%%%%%%%%%%%%%%%%%%%%%%%%%%%linha2
b^{k}\mathcal{H}_{21}^{k}   & (b^{k})^{2}\mathcal{H}_{22}^{k} &
0 & 1 \\
%%%%%%%%%%%%%%%%%%%%%%%%%%%%%%%%%%%%%%%%%%%%%%%%linha3
1 & 0 & \mathcal{H}_{33}^{k-1}  & b^{k}\mathcal{H}_{43}^{k-1}  \\
%%%%%%%%%%%%%%%%%%%%%%%%%%%%%%%%%%%%%%%%%%%%%%%%linha4
0 & 1 & b^{k}\mathcal{H}_{43}^{k-1}  &
(b^{k})^{2}\mathcal{H}_{44}^{k-1}   \\
\end{array}\right) \nonumber
\end{eqnarray}
and
\begin{eqnarray}
\mathrm{R}^{k}&=& \left(
\begin{array}{cccc}
%%%%%%%%%%%%%%%%%%%%%%%%%%%%%%%%%%%%%%%%%%%%%%%%linha1
0 & 0 & 0 \quad & 0 \\
%%%%%%%%%%%%%%%%%%%%%%%%%%%%%%%%%%%%%%%%%%%%%%%%linha2
0 & 0 & 0 \quad & 0 \\
%%%%%%%%%%%%%%%%%%%%%%%%%%%%%%%%%%%%%%%%%%%%%%%%linha3
\mathcal{H}_{13}^{k-1}-1 & b^{k-1}\mathcal{H}_{23}^{k-1} & 0 \quad & 0\\
%%%%%%%%%%%%%%%%%%%%%%%%%%%%%%%%%%%%%%%%%%%%%%%%linha4
b^{k}\mathcal{H}_{41}^{k-1} &
b^{k}b^{k-1}( \mathcal{H}_{24}^{k-1}-\mathcal{B}^{k-1}) & 0 \quad & 0\\
\end{array}\right).  \nonumber
\end{eqnarray}
The matrix $\left(\mathrm{R}^{m}\right)^T$ is the transpose of
$\mathrm{R}^{m}$ and $b^m \equiv 1/\sqrt{B^m}$.

The Gaussian integral in Eq. (\ref{eqexp2}) can be solved
immediately and the result is
\begin{eqnarray}
  \int
  \prod_{k=1}^{N-1} \left\{
  \mathrm{d} [\delta u^k] \mathrm{d} [\delta v^k]
  \mathrm{d} [\delta \tilde U^k] \mathrm{d} [\delta \tilde V^k]
    \right\}
  \exp{\left\{-\frac{1}{2} \delta  \tilde{\mathbf{x}}^T [-\delta^2
  \tilde F(\bar{\mathbf{x}})]
  \delta  \tilde{\mathbf{x}} \right\} } = \sqrt{\frac{(2\pi)^{4(N-1)}}
  {\det[-\delta^2 \tilde F(\bar{\mathbf{x}})]}}.
\label{gaussian}
\end{eqnarray}
The evaluation of the fluctuation determinant is the most lengthy
step of the whole calculation. Here we shall only briefly describe
the main steps of the calculation.

We call $\Delta^N$ the determinant of $[-\delta^2 \tilde
F(\bar{\mathbf{x}})]$ and we solve it by the Laplace method of
reducing it to smaller determinants. In this process we are lead
to define 5 auxiliary matrices whose determinants, together with
$\Delta^N$, form a closed set of six discrete recurrence
relations. The five determinants are called $\Delta^N_0$ and
$\Delta_{ij}^N$, for $i,j=1$ or $2$. The former is the determinant
of $[-\delta^2 \tilde F(\bar{\mathbf{x}})]$ without the first two
lines and columns, while $\Delta_{ij}^N$ is the determinant of
$[-\delta^2 \tilde F(\bar{\mathbf{x}})]$ without the first, second
and i-th lines and without the first, second and j-th columns.
Taking the limit of $N\rightarrow \infty$ in the mentioned set of
relations we obtain the following set of linear differential
equations:
\begin{eqnarray}
\dot{\mathbf{D}}= \frac{i}{\hbar}\left(
   \begin{array}{cccccc}
   0 & -\mathbf{H}_{22} & -\mathbf{H}_{11} & -\mathbf{H}_{21} &
   -\mathbf{H}_{21} & 0  \\
%%%%%%%%%%%%%%%%%%%%%%%%%%%%%%%%%%%%%%%%%%%%%%%%%%%%%%%%%%%%%%%%%%%%%%%%%%
   \mathbf{H}_{44} & -2\mathbf{H}_{24} & 0 & -\mathbf{H}_{41} &
   -\mathbf{H}_{41} & -\mathbf{H}_{11}  \\
%%%%%%%%%%%%%%%%%%%%%%%%%%%%%%%%%%%%%%%%%%%%%%%%%%%%%%%%%%%%%%%%%%%%%%%%%%
   \mathbf{H}_{33} & 0 & -2\mathbf{H}_{13} &-\mathbf{H}_{23} &
   -\mathbf{H}_{23} & -\mathbf{H}_{22}  \\
%%%%%%%%%%%%%%%%%%%%%%%%%%%%%%%%%%%%%%%%%%%%%%%%%%%%%%%%%%%%%%%%%%%%%%%%%%
   \mathbf{H}_{43} & -\mathbf{H}_{23} & -\mathbf{H}_{41} &-\mathbf{H}_+ &
   0 & \mathbf{H}_{21}  \\
%%%%%%%%%%%%%%%%%%%%%%%%%%%%%%%%%%%%%%%%%%%%%%%%%%%%%%%%%%%%%%%%%%%%%%%%%%
   \mathbf{H}_{43} & -\mathbf{H}_{23} & -\mathbf{H}_{41} & 0 &
   -\mathbf{H}_+ & \mathbf{H}_{21} \\
%%%%%%%%%%%%%%%%%%%%%%%%%%%%%%%%%%%%%%%%%%%%%%%%%%%%%%%%%%%%%%%%%%%%%%%%%%
   0 & \mathbf{H}_{33} & \mathbf{H}_{44} & -\mathbf{H}_{43} &
   -\mathbf{H}_{43} & -2\mathbf{H}_+
   \end{array}
\right){\mathbf{D}} \label{del}
\end{eqnarray}
where $\mathbf{D}^T(t)= (\Delta (t), \Delta_{11}(t),
\Delta_{22}(t), \Delta_{12}(t), \Delta_{21}(t), \Delta_{0}(t))$ is
a vector containing the original determinant $\Delta (t)$ and the
five auxiliary ones. $\mathbf{H}_{ij}$ are elements of the matrix
\begin{eqnarray}
\mathbf{H}\equiv
\left(\begin{array}{cccc} \frac{\partial^2
\tilde{H}}{\partial u
\partial u} & d \frac{\partial^2 \tilde{H}}{\partial u \partial U} &
\frac{\partial^2 \tilde{H}}{\partial u \partial v  }& d
\frac{\partial^2 \tilde{H}}{\partial u \partial V  }\\
d \frac{\partial^2 \tilde{H}}{\partial U \partial u}& d^2 \!
\left[ \frac{\partial^2 \tilde{H}}{\partial U \partial U}
+\frac{2V }{1+UV}\frac{\partial \tilde{H}}{\partial U}\right] & d
\frac{\partial^2 \tilde{H}}{\partial U
\partial v  } & d^2 \! \left[ \frac{\partial^2
\tilde{H}}{\partial U \partial V  } + \frac{
V\frac{\partial \tilde{H}}{\partial V  }+ U\frac{\partial \tilde{H}}{\partial U}
}{1+UV} \right]\\
\frac{\partial^2 \tilde{H}}{\partial v   \partial u} & d
\frac{\partial^2 \tilde{H}}{\partial v   \partial U} &
\frac{\partial^2 \tilde{H}}{\partial v   \partial v }& d
\frac{\partial^2 \tilde{H}}{\partial v   \partial V  }\\
d \frac{\partial^2 \tilde{H}}{\partial V   \partial u}& d^2 \!
\left[\frac{\partial^2 \tilde{H}}{\partial U
\partial V  } + \frac{V\frac{\partial \tilde{H}}{\partial V  }+ U\frac{\partial \tilde{H}}{\partial U}
  }{1+UV   }\right] & d \frac{\partial^2
\tilde{H}}{\partial V   \partial v  }& d^2 \! \left[
\frac{\partial^2 \tilde{H}}{\partial V
\partial V  } +\frac{2U}{1+UV   }\frac{\partial
\tilde{H}}{\partial V  }
\right]\\
\end{array}\right) \nonumber ,
\end{eqnarray}
calculated with the classical trajectory where $d=(1+\bar U \bar
V)/\sqrt{2j}$ and $\mathbf{H}_\pm \equiv \mathbf{H}_{13} \pm
\mathbf{H}_{24}$.

Setting $\mathbf{D}'= \mathbf{D} e^{-\frac{i}{\hbar} \int
\mathbf{H}_+ dt}$ we obtain the more symmetric form
\begin{eqnarray}
\dot{\mathbf{D}}'= \frac{i}{\hbar} \cal{A} \, {\mathbf{D}}'
\label{dd}
\end{eqnarray}
with
\begin{eqnarray}
\cal{A}=  \left(
   \begin{array}{cccccc}
   \mathbf{H}_+ & -\mathbf{H}_{22} & -\mathbf{H}_{11} & -\mathbf{H}_{21}
   & -\mathbf{H}_{21} & 0  \\
%%%%%%%%%%%%%%%%%%%%%%%%%%%%%%%%%%%%%%%%%%%%%%%%%%%%%%%%%%%%%%%%%%%%%%%%%%
   \mathbf{H}_{44} & \mathbf{H}_{-} & 0 & -\mathbf{H}_{41}
   & -\mathbf{H}_{41} & -\mathbf{H}_{11}  \\
%%%%%%%%%%%%%%%%%%%%%%%%%%%%%%%%%%%%%%%%%%%%%%%%%%%%%%%%%%%%%%%%%%%%%%%%%%
   \mathbf{H}_{33} & 0 & -\mathbf{H}_{-} &-\mathbf{H}_{23}
   & -\mathbf{H}_{23} & -\mathbf{H}_{22}  \\
%%%%%%%%%%%%%%%%%%%%%%%%%%%%%%%%%%%%%%%%%%%%%%%%%%%%%%%%%%%%%%%%%%%%%%%%%%
   \mathbf{H}_{43} & -\mathbf{H}_{23} & -\mathbf{H}_{41} &0 & 0
   & \mathbf{H}_{21}  \\
%%%%%%%%%%%%%%%%%%%%%%%%%%%%%%%%%%%%%%%%%%%%%%%%%%%%%%%%%%%%%%%%%%%%%%%%%%
   \mathbf{H}_{43} & -\mathbf{H}_{23} & -\mathbf{H}_{41} & 0 & 0
   & \mathbf{H}_{21} \\
%%%%%%%%%%%%%%%%%%%%%%%%%%%%%%%%%%%%%%%%%%%%%%%%%%%%%%%%%%%%%%%%%%%%%%%%%%
   0 & \mathbf{H}_{33} & \mathbf{H}_{44} & -\mathbf{H}_{43}
   & -\mathbf{H}_{43} & -\mathbf{H}_+
   \end{array}
\right).  \label{amat}
\end{eqnarray}

The equations for ${\mathbf{D}}'$ are intimately related to the
linearized equations of motion around the critical trajectory. To
see this we go back to our notation as is Eqs.~ (\ref{compactn}).
Setting the small displacements
$\delta\chi_i(t)\equiv\chi_i(t)-\bar\chi_i(t)$ around the critical
trajectory and defining
\begin{eqnarray}
\xi_1 (t)= \delta u(t), \qquad \xi_2 (t)=
\frac{\sqrt{2j}}{(1+U(t)V(t))} \delta U(t), \\ \nonumber\\
\xi_3 (t)= \delta v(t), \qquad \xi_4 (t)=
\frac{\sqrt{2j}}{(1+U(t)V(t))} \delta V(t),
\end{eqnarray}
we can construct an anti-symmetric tensor $T_{ik}(t)= \xi_i
(t)\xi_k' (t)- \xi_i'(t) \xi_k(t)$, where $\xi_i(t)$ and
$\xi_i'(t)$ are independent displacements from the critical
trajectory. The tensor $T$ has six independent components, which
can be arranged into a new vector defined by ${\cal T}^T = ({T}_{34}(t),
{T}_{23}(t), {T}_{41}(t), {T}_{13}(t), {T}_{42}(t), {T}_{12}(t))$
whose equation of motion is exactly $\dot{{\cal T}}=i {\cal
A} {\cal T}/\hbar$. Putting things together we find that
\begin{eqnarray}
\det[-\delta^2\tilde F (\bar{\mathbf{x}})]\equiv\Delta (T)= T_{34}
(T)\exp \left\{ -\frac{i}{\hbar} \int_0^T \mathbf{H}_+ dt \right\}.
\end{eqnarray}

Since $T_{34}$ is related to the linearized motion around the
critical trajectory, it can be easily written in terms of the
tangent matrix or in terms of second derivatives of the action.
Working out the details we find
\begin{eqnarray}
\Delta (T) = \frac{(1+U(0)V(0))}{(1+U(T)V(T))} \left[ \det
\mathrm{M}_{bb} \right] e^{ -\frac{i}{\hbar} \int_0^T \mathbf{H}_+
dt } , \label{eq103}
\end{eqnarray}
where $M_{bb}$ is the lower right 2 by 2 block of the tangent
matrix in the coordinates $\chi_i$:
\begin{eqnarray}
\mathrm{M}(T)= \left(
\begin{array}{cccc}
\mathrm{M}_{11} (T) &\mathrm{M}_{12} (T)
&\mathrm{M}_{13} (T)&\mathrm{M}_{14}(T) \\
\mathrm{M}_{21} (T) &\mathrm{M}_{22}(T)
&\mathrm{M}_{23}(T) &\mathrm{M}_{24} (T)\\
\mathrm{M}_{31} (T) &\mathrm{M}_{32} (T)
&\mathrm{M}_{33} (T)&\mathrm{M}_{34} (T)\\
\mathrm{M}_{41} (T) &\mathrm{M}_{42} (T)
&\mathrm{M}_{43} (T)&\mathrm{M}_{44} (T)\\
\end{array}
\right) \equiv \left(
\begin{array}{ll}
\mathrm{M}_{aa} &\mathrm{M}_{ab} \\
\mathrm{M}_{ba} &\mathrm{M}_{bb} \\
\end{array}
\right). \label{mt}
\end{eqnarray}
Differentiating both sides of Eqs. (\ref{dif1}) and (\ref{dif2})
and conveniently re-arranging the terms we identify

\begin{eqnarray}
\mathrm{M}_{bb} =
\frac{i\hbar}
{\frac{\partial^2\mathcal{S}}{\partial u'\partial v''}
\frac{\partial^2\mathcal{S}}{\partial U'\partial V''}-
\frac{\partial^2\mathcal{S}}{\partial u'\partial V''}
\frac{\partial^2\mathcal{S}}{\partial U'\partial v''}}
\left(
\begin{array}{cc}
-\frac{\partial^2\mathcal{S}}{\partial U'\partial V''}&
2j \left(\frac{1}{1+U'V'}\right)^2\frac{\partial^2\mathcal{S}}{\partial u'\partial V''}\\
\frac{\partial^2\mathcal{S}}{\partial U'\partial v''}&
-2j \left(\frac{1}{1+U'V'}\right)^2\frac{\partial^2\mathcal{S}}{\partial u'\partial v''}
\end{array}\right). \label{MtoS2}
\end{eqnarray}

%%%%%%%%%%%%%%%%%%%%%%%%%%%%%%%%%%%%%%%%%%%%%%%%%%%%%%
\subsection{The final formula}

Collecting all the results for the exponent and pre-factor, the
final formula for the semiclassical limit of the canonical-spin
coherent state propagator becomes
\begin{eqnarray}
K \left( z''^*,  s''^*, z', s', T \right)
  & = &
 \left[\left(\frac{1+ U''  V''}
             {1+ U' V'} \right)
 \frac{1}{\det \mathrm{M}_{bb}}\right]^{1/2} \exp{\left\{
   \frac{i}{\hbar} (\mathcal{S} + \mathcal{G})
   - \Lambda \right\} }  \label{PP}
\end{eqnarray}
where
\begin{eqnarray}
\mathcal{S}(z''^*,s''^*,z',s',T) &=&
     \int_0^T
     \left\{
     \frac{i\hbar}{2} \left(\dot{u} v -
     \dot{v} u \right) -i\hbar j
     \left(  \frac{U \dot{V} -
     V \dot{U}}
     { 1 + U V }  \right)  - {\tilde H}
     \right\}  dt  \nonumber \\
        & -&\frac{i\hbar}{2}\left(
        u' v' + u'' v'' \right)
   - i\hbar j \ln \left[
   (1+U' V')(1 + U'' V'') \right],
   \label{S}
\\ \nonumber \\
\mathcal{G}(z''^*, s''^*,z',s',T)
  &=& \displaystyle{\frac{1}{2}\int_0^T \left\{
    \frac{\partial^2 \tilde{H}}{\partial v \partial u}+
    \frac{1}{2}
    \left[ \frac{\partial}{\partial V}
    \frac{(1+ V U)^2}{2j}
    \frac{\partial \tilde{H}}{\partial U}+
    \frac{\partial}{\partial U}
    \frac{(1+ V U)^2}{2j}
    \frac{\partial \tilde{H}}{\partial V}
    \right]
    \right\}dt} \nonumber \\
    &\equiv& \frac{1}{2}\int_0^T
\mathbf{H}_+ dt
    . \label{G}
\end{eqnarray}
and
\begin{equation}
\Lambda = \frac{1}{2}\left( |z'|^2 + |z''|^2\right) + j \ln
\left[(1+|s'|^2)(1+|s''|^2) \right] .
\end{equation}

Alternatively, the pre-factor can be written explicitly in terms
of derivatives of the action acording to Eq. (\ref{MtoS2}),
\begin{eqnarray}
 \left[\left(\frac{1+ U''  V''}
             {1+ U' V'} \right)
 \frac{1}{\det \mathrm{M}_{bb}}\right]^{1/2}
 \rightarrow
  \left[\frac{(1+ U'' V'')(1+ U' V')}{2j}
  {[\det \Sigma]} \right]^{1/2}  \label{PP2},
\end{eqnarray}
where
\begin{eqnarray}
\Sigma = \frac{i}{\hbar} \left(
\begin{array}{cc}
\frac{\partial^2 \mathcal{S}}{\partial  u'\partial  v''} &
\frac{\partial^2 \mathcal{S}}{\partial  u'\partial  V''} \\
\frac{\partial^2 \mathcal{S}}{\partial  U'\partial  v''} &
\frac{\partial^2 \mathcal{S}}{\partial  U'\partial  V''}
\end{array}
\right).
\end{eqnarray}
All quantities are to be calculated at the stationary trajectory
(we have removed the bar on top these quantities to simplify the
notation).

%%%%%%%%%%%%%%%%%%%%%%%%%%%%%%%%%%%%%%%%%%%%%%%%
%%%%%%%%%%%%%%%%%%%%%%%%%%%%%%%%%%%%%%%%%%%%%%%%
\section{Simple Applications}

In this section we shall apply the semiclassical formula
Eq.(\ref{PP}) to three simple situations: (a) non-interacting spin
and field operators; (b) the limit of very large spins and; (c)
the case of spin $1/2$. In each case we shall see that the
propagator obtained from Eq.(\ref{PP}) reduces to well known
results.

\subsection{Non-interacting Hamiltonian}

If the Hamiltonian can be separated into $\hat{H} = \hat{H}_z +
\hat{H}_s$, where $\hat{H}_z=\hat{H}_z(\hat a,\hat a^\dagger)$ and
$\hat{H}_s= \hat{H}_s(\hat{J}_+, \hat{J}_-, \hat{J}_z)$, then
$\tilde H \equiv \langle z,s |\hat{H}|z,s\rangle = \tilde{H}_z +
\tilde{H}_s$, where $\tilde{H}_z \equiv \langle z
|\hat{H}_z|z\rangle = \tilde{H}_z(z^*,z)$ and $\tilde{H}_s \equiv
\langle s |\hat{H}_s|s\rangle=\tilde{H}_s(s^*,s)$. Therefore, the
complex action of Eq. (\ref{action}) takes the form
\begin{eqnarray}
\mathcal{S}(z''^*,s''^*,z',s',T) =
\mathcal{S}_z(z''^*,z',T)+\mathcal{S}_s(s''^*,s',T) ,
\end{eqnarray}
where
\begin{eqnarray}
\begin{array}{l}
     \mathcal{S}_z(z''^*,z',T)=
     \int_0^T
     \left\{
     \frac{i\hbar}{2} \left(\dot{u} v - \dot{v} u \right)
     - {\tilde H}_z
     \right\}  dt -
     \frac{i\hbar}{2}\left( u'v'+u''v'' \right),
     \\ \\
     \mathcal{S}_s(s''^*,s',T)=
     \int_0^T
     \left\{
     -i\hbar j
     \left(  \frac{U \dot{V} - V \dot{U}}
     { 1 + UV }  \right)  - {\tilde H}_s
     \right\}  dt
   -i\hbar j \ln \left[
   (1+U'V')(1+U''V'') \right]   .
\end{array}
\end{eqnarray}
Moreover, the term $\mathcal G$ of Eq. (\ref{G}) becomes
\begin{eqnarray}
\mathcal{G}(z''^*,s''^*,z',s',T) = \mathcal{G}_z(z''^*,z',T)+ \mathcal{G}_s(s''^*,s',T),
\end{eqnarray}
where
\begin{eqnarray}
\begin{array}{l}
    \mathcal{G}_z(z''^*,z',T)=
    \frac{1}{2}\int_0^T
    \frac{\partial^2 \tilde{H}}{\partial v \partial u}
    dt    ,\\ \\
    \mathcal{G}_s(s''^*,s',T)=
    \frac{1}{4}\int_0^T
    \left[ \frac{\partial}{\partial V}
    \frac{(1+UV)^2}{2j}
    \frac{\partial \tilde{H}}{\partial U}+
    \frac{\partial}{\partial U}
    \frac{(1+UV)^2}{2j}
    \frac{\partial \tilde{H}}{\partial V}
    \right]
    dt    .
\end{array}
\end{eqnarray}
Finally, the $\det \Sigma$ simplifies to
\begin{eqnarray}
\det \Sigma = - \frac{1}{\hbar^2} \left[ \frac{\partial^2
\mathcal{S}_z}{\partial u'\partial v''} \frac{\partial^2
\mathcal{S}_s}{\partial U'\partial V''}\right].
\end{eqnarray}

Therefore, for non-interacting systems, Eq. (\ref{PP}) amounts to
\begin{eqnarray}
K \left( z''^*,  s''^*, z', s', T \right)
   &\equiv& K_z\left( z''^*,  z', T \right) \times K_s \left( s''^*,  s', T
   \right),
\end{eqnarray}
where
\begin{eqnarray}
K_z \left( z''^*, z', T \right)
  &=&
  \sqrt{\frac{i}{\hbar}
    \frac{\partial^2 \mathcal{S}_z}{\partial u'\partial v''}}
  ~e^{
   \frac{i}{\hbar} (\mathcal{S}_z
   +    \mathcal{G}_z )+ \frac{1}{2}|z'|^2 + \frac{1}{2}|z''|^2} ,\nonumber
   \\ \nonumber \\
K_s \left( s''^*,  s', T \right) &=&
    \sqrt{\frac{i}{\hbar}\frac{(1+U''V'')(1+U'V')}{2j}
    \frac{\partial^2 \mathcal{S}_s}{\partial U'\partial V''}}
  ~e^{
   \frac{i}{\hbar} (\mathcal{S}_s
   +   \mathcal{G}_s) -j \ln
\left[(1+|s'|^2)(1+|s''|^2) \right]
   }
\end{eqnarray}
are exactly the semiclassical propagators known in the literature
for the Weyl and SU(2) groups respectvely (see, for example, Refs.
\cite{bar} and \cite{stone}).

%%%%%%%%%%%%%%%%%%%%%%%%%%%%%%%%%%%%%%%%%%%%%%%%%%%%%%
\subsection{The limit of large spin}

Following Perelomov \cite{perelomov}, we set  $s=w/\sqrt{2j}$,
$\hat{J}_+=\sqrt{2j}\hat a^{\dagger}$ and let $j \rightarrow
\infty$. In this limit the spin coherent states transform into
canonical coherent states:
\begin{eqnarray}
|s\rangle
%\equiv \frac{\exp \left\{ s \hat{J}_+\right\}}{1+|s|^2}|-j\rangle
\rightarrow
|w\rangle = \frac{ \exp \left\{ w \hat{a}^{\dagger} \right\}} {\left(1+\frac{|w|^2/2}{j}\right)^j}|-j\rangle
\approx e^{ w\hat{a}^{\dagger}-\frac{1}{2}|w|^2}|0\rangle .
\end{eqnarray}
In this case, discarding terms smaller than $j^{-1}$ we obtain

\begin{eqnarray}
&&j\frac{s\dot{s}^* - \dot{s}s^*}{1+s s^*}
%= \frac{1}{2} \frac{w\dot{w}^*-\dot{w}w^*}{1+\frac{ww^*}{2j}}
\rightarrow \frac{1}{2} (w\dot{w}^*-\dot{w}w^*) , \nonumber\\
&&j\ln \left[ (1+s's'^*)(1+s''s''^*)\right]
\rightarrow -\frac{1}{2} \left(w'w'^*+w''w''^*\right), \label{eq127} \\
&&\frac{\partial}{\partial s^*} \frac{(1+ss^*)^2}{2j}
\frac{\partial \tilde H}{\partial s}+ \frac{\partial}{\partial s}
\frac{(1+ss^*)^2}{2j} \frac{\partial \tilde H}{\partial s^*}
\rightarrow 2\frac{\partial^2 \tilde H}{\partial w  \partial
w^*}\nonumber
\end{eqnarray}
and
\begin{eqnarray}
\begin{array}{l}
\frac{(1+s''s''^*)(1+s's'^*)}{2j} \det
\left(
\begin{array}{cc}
\frac{i}{\hbar}\frac{\partial^2 \mathcal{S}}{\partial z'\partial z''^*} &
\frac{i}{\hbar}\frac{\partial^2 \mathcal{S}}{\partial z'\partial s''^*} \\
\frac{i}{\hbar}\frac{\partial^2 \mathcal{S}}{\partial s'\partial z''^*} &
\frac{i}{\hbar}\frac{\partial^2 \mathcal{S}}{\partial s'\partial s''^*}
\end{array}
\right)  \rightarrow \det  \left(
\begin{array}{cc}
\frac{i}{\hbar}\frac{\partial^2 \mathcal{S}}{\partial z'\partial z''^*} &
\frac{i}{\hbar}\frac{\partial^2 \mathcal{S}}{\partial z'\partial s''^*} \\
\frac{i}{\hbar}\frac{\partial^2 \mathcal{S}}{\partial w'\partial z''^*} &
\frac{i}{\hbar}\frac{\partial^2 \mathcal{S}}{\partial w'\partial w''^*}
\end{array}
\right) \end{array}. \label{eq130}
\end{eqnarray}
Equations (\ref{eq127}) to (\ref{eq130}) applied to Eqs.
(\ref{PP}) and (\ref{PP2}) produces the two-dimensional
semiclassical propagator for canonical coherent states
\cite{rib1}.

%%%%%%%%%%%%%%%%%%%%%%%%%%%%%%%%%%%%%%%%%%%%%%%%%%%%%%
\subsection{Spin-$\frac{1}{2}$ Systems}

The semiclassical approximation developed in sections II to IV
employed explicitly the limit $j \rightarrow \infty$. In this
subsection we discuss the application of Eq. (\ref{PP}) to
spin-$\frac{1}{2}$ systems, whose general Hamiltonian is given by
\begin{eqnarray}
\hat{H}=\hat{H}_0+\hat{H}_s \equiv \hat{H}_0(\hat a,\hat
a^\dagger) + \hbar \hat{\mathbf s} \cdot \hat{\mathbf C}(\hat
a,\hat a^\dagger).
\end{eqnarray}
In this case the classical Hamiltonian reads
\begin{eqnarray}
\tilde H (u,v,U,V)&=& \langle z|\tilde{H}_0 |z\rangle + \hbar
\langle s| \hat{\mathbf s} |s\rangle \cdot \langle z|\hat{\mathbf
C} |z\rangle = \tilde{H}_0 (u,v) + \frac{\hbar}{2} \tilde{H}_s
(u,v,U,V),
\end{eqnarray}
where
\begin{eqnarray}
\tilde{H}_s (u,v,U,V)= {C}_1(u,v) \frac{U+V}{1+UV} -i {C}_2(u,v)
\frac{V-U}{1+UV} - {C}_3 (u,v)\frac{1-UV}{1+UV} .
\end{eqnarray}
and $\langle z|\hat{\mathbf C} |z\rangle \equiv ({C}_1(u,v),
{C}_2(u,v), {C}_3(u,v))$.

The equations of motion are given explicitly by
\begin{eqnarray}
\begin{array}{lll}
\dot{v}&=& \frac{i}{\hbar}  \frac{\partial }{\partial u}
(\tilde{H}_0+ \frac{\hbar}{2} \tilde{H}_s), \\
\dot{u}&=&-\frac{i}{\hbar} \frac{\partial }{\partial v}
(\tilde{H}_0+ \frac{\hbar}{2} \tilde{H}_s), \\  \dot{V}&=&
\frac{i}{2} \left[ (C_1+iC_2)- (C_1-iC_2) V^2 + 2C_3 V  \right],
\\ \dot{U}&=& \frac{i}{2} \left[ (C_1+iC_2) U^2 -(C_1-iC_2) - 2C_3
U\right].
\end{array}
\label{eqm}
\end{eqnarray}

In the limit of small $\hbar$ we can drop the terms
$\frac{\hbar}{2} \tilde{H}_s$ on the first two equations and
decouple $u$ and $v$ from the spin variables $U$ and $V$. These,
on the other hand, describe the precession of the spin in the
external field ${\bf C}(u,v)$ generated by the orbital motion. In
this approximation the orbital part of the action also separates
from the total action and the semiclassical propagator can be
written as
\begin{equation}
K \left( z''^*,  s''^*, z', s', T \right)
  = K_z\left( z''^*,  z', T \right) \times {K_s}_{[u,v]}
   \left( s''^*,  s', T \right), \label{prophalf}
\end{equation}
where $K_z$ is the one-dimensional canonical propagator and
${K_s}_{[u,v]}$ can be written as \cite{kochetov}
\begin{equation}
{K_s}_{[u,v]} \left( s''^*,  s', T \right) = \frac{a^*(t) -
b^*(t)s' + b(t)s''^* + a(t) s''^* s'}{(1+|s''|^2)(1+|s'|^2)} .
\label{propspinhalf}
\end{equation}
The coefficients $a(t)$ and $b(t)$ are obtained from the
differential equation
\begin{equation}
\frac{{\rm d}W}{{\rm d}t} = -\frac{i}{2} \mathbf{\sigma} \cdot
{\bf C} (t) W(t)
\end{equation}
where
\begin{equation}
W(t) = \left( \begin{array}{ll} a(t) & b(t) \\
-b^*(t) & a^*(t) \end{array} \right),
\end{equation}
$\mathbf{\sigma}$ are the Pauli matrices and $W(0)=\mathbf{1}$.
Since Eq.(\ref{propspinhalf}) is the exact propagator for a spin
in an external field, Eq.(\ref{prophalf}) can also be derived
directly from the path integral approach
\begin{equation}
K \left( z''^*,  s''^*, z', s', T \right)
  = \int \frac{{\cal D}[u] {\cal D}[v]}{\pi}
     {K_s}_{[u,v]} \left( s''^*,  s', T \right)
     e^{\frac{i}{\hbar} F_{z0}[u,v,T] }
\end{equation}
where the steepest descent approximation is performed only in the
orbital action $F_{z0}$ (which contains only $H_0$). The spin
propagator $K_s$ is viewed as a slow varying pre-factor and is
simply calculated at the stationary trajectory
\cite{zaitsev1,zaitsev2}. This shows that the semiclassical
formula Eq. (\ref{PP}) can also be used for systems with spin
$j=1/2$, in spite of the large spin limit considered in its
derivation.\\

\noindent Acknowledgments

\noindent This work was partly supported by FAPESP and CNPq. ADR
and MAMA thank Dr. Marcel Novaes and Dr. Fernando Parisio for
helpful discussions.

%%%%%%%%%%%%%%%%%%%%%%%%%%%%%%%%%%%%%%%%%%%%%%%%%%%%%%


\begin{thebibliography}{20}

\bibitem{feynman} R.~P. Feynman and A.~R. Hibbs, {\em Quantum Mechanics
and Path Integrals}, McGraw-Hill, New York, 1965.

\bibitem{books} A.M. Ozorio de Almeida, {\it Hamiltonian Systems:
Chaos and Quantization}, Cambridge University Press (1989);
M.~C.~Gutzwiller, {\it Chaos in classical and quantum physics},
Springer, New York, (1990); H.J. Stockmann, {\it Quantum Chaos: An
Introduction}, Cambridge University Press, (1999).

\bibitem{klauder78} J.~R. Klauder,  {\em Continuous Representations and Path Integrals, Revisited}, in G.~J. {Papa\-do\-pou\-los} and J.~T. Devreese, editors, {\em Path Integrals}, NATO Advanced Study Institute, Series B: Physics, New York, 1978.
Plenum.

\bibitem{klauder79} J.~R. Klauder, Phys. Rev. D {\bf 19}, 2349
(1979).

\bibitem{weis82} Y.~Weissman, J. Chem. Phys. {\bf 76}, 4067
(1982).

\bibitem{weis83} Y.~Weissman, J. Phys. A: Math. and Gen. {\bf 16}, 2693
(1983).

\bibitem{miller} W.~H. Miller, Adv. Chem. Phys. {\bf 25}, 69
(1974).

\bibitem{adachi} S. Adachi, Ann. of Phys. {\bf 195}, 45 (1989).

\bibitem{bar0} M. Baranger and M.A.M. de Aguiar, (1989) unpublished
notes.

\bibitem{xav3} A. L. Xavier Jr. and M. A. M. de Aguiar, Phys. Rev. Lett.
{\bf 79}, 3323 (1997).

\bibitem{xav1} A. L. Xavier Jr. and M. A. M. de Aguiar, Ann. of Phys.
{\bf 252}, 458 (1996).

\bibitem{xav2} A. L. Xavier Jr. and M. A. M. de Aguiar, Phys. Rev. A
{\bf 54}, 1808 (1996).

\bibitem{kochetov98} E.~A. Kochetov, J. Phys. A
{\bf 31}, 4473 (1998).

\bibitem{bar} M.~Baranger, M. A. M. de Aguiar, F. Keck, H. J. Korsch
and B.~Schellaa\ss, J. Phys. A {\bf 34}, 7227 (2001).

\bibitem{rib1} A.~D.~Ribeiro, M. A. M. de Aguiar and M. Baranger,
Phys. Rev. E {\bf 69}, 066204 (2004).

\bibitem{coe} L.C. dos Santos and M.A.M. de Aguiar, Braz. J. Phys
{\bf 35} 175 (2005).

\bibitem{aguiarjpa} M. A. M. de Aguiar, M.~Baranger, L. Jaubert,
Fernando Parisio and A.~D. Ribeiro, J. Phys. A {\bf 38}, 4645 (2005).

\bibitem{rubin} A.~Rubin and J.~R. Klauder, Ann. of Phys.,
{\bf 241} 212 (1995).

\bibitem{rib3} A.~D. Ribeiro, M. Novaes and M. A. M. de Aguiar,
Phys. Rev. Lett {\bf 95}, 050405 (2005).

\bibitem{kura1} H. Kuratsuji and T. Suzuki, J. Math. Phys.
{\bf 21}, 472 (1979).

\bibitem{sol} H. Solari, J. Math. Phys. {\bf 28}, 1097 (1987).

\bibitem{vieira} V.~R. Vieira and P.~D. Sacramento,
Nucl. Phys. B {\bf 448}, 331 (1995).

\bibitem{kochetov} E.~A. Kochetov, J. Math. Phys.
{\bf 36}, 4667 (1995).

\bibitem{stone} M. Stone, K.~S. Park and A. Garg, J. Math. Phys.
{\bf 41}, 8025 (2000).

\bibitem{plet} M. Pletyukhov, J. Math. Phys. {\bf 45}, 1859 (2004).

\bibitem{garg} A. Garg, E. Kochetov, K.~S. Park and M. Stone,
J. Math. Phys. {\bf 44}, 48 (2003).

\bibitem{marcel1} M. Novaes, quant-ph/0505224v1.

\bibitem{alscher} A. Alscher and H. Grabert,
Eur. Phys. J. D {\bf 14}, 127 (2001).

\bibitem{plet2002} M. Pletyukhov, Ch. Amann, M. Mehta and M. Brack,
Phys. Rev. Lett. {\bf 89}, 116601 (2002).

\bibitem{plet2003} M. Pletyukhov, O. Zaitsev,
J. Phys. A {\bf 36}, 5181 (2003).

\bibitem{amann} Ch. Amann and M. Brack,
J. Phys. A {\bf 35}, 6009 (2002).

\bibitem{zaitsev1} O. Zaitsev, Diego Frustaglia and Klaus Richter,
Phys. Rev. Lett. {\bf 94}, 026809 (2005).

\bibitem{zaitsev2} O. Zaitsev, Diego Frustaglia and Klaus Richter,
cond-mat/0506171.

\bibitem{perelomov} A. Perelomov, {\em Generalized Coherent States and
their Applications} (Springer-Verlag, Berlin, 1986).

\end{thebibliography}
\end{document}